\documentstyle[aps,amssymb,psfig]{revtex}
\begin{document}

\author{Simonetta Frittelli$^{a,b}$ \and  
	Thomas P. Kling$^{b}$ \and 
	Ezra T. Newman$^{b}$ \\ 
	$^{a}$Department of Physics, Duquesne University, 
		Pittsburgh, PA 15282\\
	$^{b}$Department of Physics and Astronomy, 
		University of Pittsburgh,
		Pittsburgh, PA 15260} 
\title{
\rightline{\small{\em To appear in Phys. Rev. D (January 2001)\/}}
Image distortion in non-perturbative gravitational 
	lensing}
\date{\today} 
\maketitle

\begin{abstract}

We introduce the idea of {\it shape parameters\/} to describe the shape
of the pencil of  rays connecting an observer with a source lying on
his past lightcone.   On the basis of these shape parameters, we
discuss a  setting of image distortion in a generic (exact) spacetime,
in the form of three {\it distortion parameters\/}.   The fundamental
tool in our discussion is the  use of geodesic deviation
fields along a null geodesic to study how source shapes are propagated
and distorted on the path to an observer.   We illustrate this
non-perturbative treatment of image distortion in the case of lensing
by a Schwarzschild black hole. We conclude by showing that there is a
non-perturbative generalization of the use of Fermat's principle in
lensing in the thin-lens approximation. 

\end{abstract}

\section{Introduction}


Before we describe the ideas and content of this work, we wish to give a
brief discussion of certain ideas concerning lensing that are really
quite obvious but are often overlooked and or confused. It is absolutely
clear that any source of light emits many rays that travel in many
directions and get distorted by all the different available lenses. But
for an observer, of all those different rays, the only ones of relevance
are the rays that enter the observers ``eyes;'' all the others are
``lost'' to that observer and play no role at all in his or her
treatment of lensing. The conclusion from this is that at any one time,
it is the past lightcone of the observer that is the object to be
studied and understood. It is the bending and distortions of the rays
and pencils of rays lying on this past cone that constitute the basic
elements that make up lensing theory. When this past cone intersects a
source, that source is then seen by the observer; if it is a point
source then ideally only one ray reaches the observer, if it is a small
but finite source then a pencil of rays, lying on the past cone, reaches
the observer. In this later case the pencil can be described by the
geodesic deviation vectors (or Jacobi field) associated with just one
ray. The shape of the pencil can and will, in general, undergo changes
due to the presence of gravitational field along the path of the one
ray. It is this change in the shape of the pencil, between the source
and the observer, that constitutes image distortion. This is the point
of view that we adopt throughout this paper -- though the details are
very different, the point of view concerning the essential role of the
past cone applies equally well to the thin lens approximation (the
standard approach) and to the non-perturbative approach to lensing.

It is often the case in gravitational lensing that the distortion of
extended images provides as much information to the practicing
astrophysicist as the location of the images themselves~\cite{Tyson}.
In previous work~\cite{EFNspacetime,FNuniversal,FKNschw}, the theory of
gravitational lensing has been developed and studied from a first
principles point of view, where lensing is considered
non-perturbatively with no restrictions on the strength of the
gravitational field. This work is non-perturbative in the sense that
the key equations of lensing theory, the lens and time of arrival
equations, are derived by integrating (as far as possible) the null
geodesic equations of an (in principle) exact metric. This point of
view is in contrast to the more standard work in lensing based on the
thin lens approximation where lenses ``live'' in some fixed background
spacetime. In the present paper and its companion~\cite{FKNII}, we turn our
attention to the non-perturbative study of distortion of images due to
lensing.

Images are distorted by a lens in multiple ways. First, an image may be
magnified or demagnified, relative to what the image would look like
``in the absence of the lens.'' Magnification is due to the focusing of
the lightcone by the spacetime curvature, and the study of
magnification is closely related to catastrophe optics and singularity
theory~\cite{petters93,petters97,yellowarnold,greenarnold,berry,berry2}.
The magnification of images has proved to be an extremely useful
property of lensing systems because it often allows for the observation
of very distant sources which may not have been observed without the
magnification.

Perhaps more importantly, an extended image may be stretched along some
axis, rotated, or made generally larger or smaller relative to an
``unlensed image.'' Colloquially, we say that the stretching along an
axis and rotation of an image is related to the ``shearing'' of the
pencil of light and that an overall change in size is due to the
``convergence'' (or alternatively, ``expansion'' or ``divergence'') of
the pencil. These effects are very important in lens modeling, as the
distortion of an extended image into an arc or of identifiable parts of
an image (such as the radio jets of a QSO) provides very useful
observational constraints on the lens model.

In general relativity, the shear and the convergence are precisely
defined quantities~\cite{penrosespinors} which have values at each
point along a geodesic. These quantities, often refered to as the
optical scalars, have been widely studied in the relativity community.
Hence, the general use of these terms in lensing, where an image --
observed here -- of a source -- over there -- is said to be ``sheared''
can be a little misleading.

One goal of this paper and its companion is to propose a unified
framework for the non-perturbative discussion of the distortion of
images which can be applied in strong-field regimes. In this paper, we
introduce three quantities, the {\it shape parameters\/}, describing
the generic appearance of an elliptical image or source and show how
the shape parameters are computed from connecting vectors. We also
introduce three {\it distortion parameters\/} which correspond to
changes in the shape parameters between the source and
observer. In the following paper, we relate the shape and distortion
parameters to the optical scalars.

This paper begins by reviewing the standard (thin-lens) approach to the
distortion of images as presented in~\cite{EFS}. In Sec.~II, we introduce the
key idea of using the inverse Jacobian of the lens equations to map source
curves (the outline or boundary of sources) into image curves. We define the
shape parameters at the end of this section and briefly mention the quantities
which are referred to as the ``shear'' and ``convergence'' in the thin lens
approximation. (The full discussion of these quantities is left to the second
paper, which is devoted to the relationship between the shape parameters and
optical scalars.) In the third section of this paper, we present the
non-perturbative version of the shape parameters and define the distortion of
images. This section relies heavily on the use of connecting vectors or Jacobi
fields (we follow the lead of other authors in the  use of Jacobi fields in
lensing, in particular~\cite{penroseoptical,blandfordjacobi}). The distortion
of images in Schwarzschild spacetime is discussed at the conclusion of this
section. Finally, we provide a different perspective on the subject in
Sec.~\ref{sec:gauge} which we feel clarifies several issues regarding the role
of the choice of coordinates.

An important caveat must be stated. In a generic lensing situation, if
the past lightcone is continued arbitrarily far into the past,
caustics will always be found. We, in this work, are making the
assumption that our source that is being imaged {\it does not\/}
intersect a caustic curve. The morphology of image construction for
sources which lie across a caustic is well understood but relatively
complicated. The issue of distortion of images in this case is
something for further study.

\section{Image distortion via the thin lens approximation}

\begin{figure}
\centerline{\psfig{figure=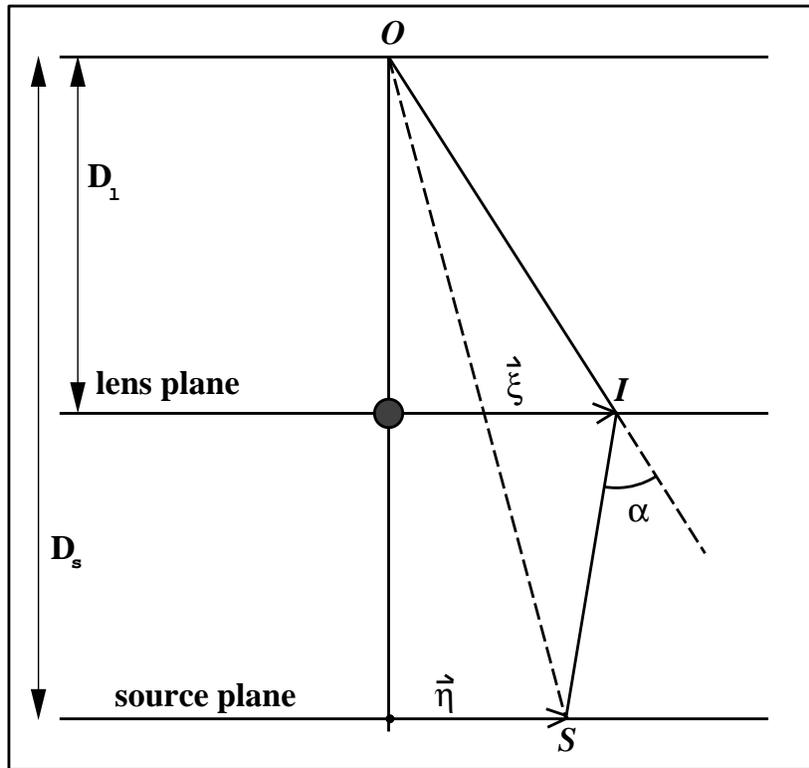,height=4in,angle=0} }

\caption{The lensing problem in standard gravitational lensing.  The
point $O$ represents the observer.  The point $S$ represents the source.
The lens mapping can be interpreted as a map that takes the point $I$ in
the lens plane into the point $S$ in the source plane.}

\label{fig:distortionI1}
\end{figure}

In this section we summarize the approach to image distortion in
standard gravitational lensing according to~\cite{EFS}. The assumptions
of the thin lens approximation are that the source, lens, and observer
are at rest in the background spacetime, that the gravitational fields
are weak, that all angles are small, and that the mass distribution of
the entire lens can be collapsed into a single two-dimensional ``lens
plane'' and is described by a surface mass-density, $\kappa
(\vec{\xi})$. As is shown in Fig.~\ref{fig:distortionI1}, light rays
are assumed to travel along the null geodesics of the background
spacetime from a point in the source plane at $S$ to a point in the
lens plane (the plane containing the lens and perpendicular to the
optical axis) at $I$. At $I$, the ray is instantaneously bent by a
bending angle, $\alpha $, and travels to the observer $O$ along another
background geodesic. A goal of the thin lens approximation is to find a
set of equations, called lens equations, which represent the path of
the light ray as a mapping between the lens plane and the source plane.

The lens plane is described by the two dimensional vector $\vec{\xi}$,
and the source plane is described by the two dimensional vector
$\vec{\eta}$. It is convenient to introduce two scaled vectors

\begin{equation}
	\vec{x}
   =	\frac{\vec{\xi}}{D_l},
	\quad \quad \text{and}\quad \quad 
	\vec{y}
   =	\frac{\vec{\eta}}{D_s},  
\label{scaled}
\end{equation}

\noindent which describe the lens $(\vec{x})$ and source $(\vec{y})$
planes. Here $D_s$ and $D_l$ are the angular-diameter distances
between the observer and source and observer and lens.

The dynamics of the thin lens approximation is introduced through a
deflection potential

\begin{equation}
	\psi(\vec{x}) 
   = 	\frac{1}{\pi} 
	\int_{\mbox{R}^2} \, dx'^2 
	\kappa (\vec{x}') \ln|\vec{x}-\vec{x}'|,  
\label{deflectionpot}
\end{equation}

\noindent where $\kappa(\vec{x})$ is the surface mass-density,
satisfying the two-dimensional Laplace equation

\begin{equation}
	\nabla^2\psi(\vec{x}) = 2\kappa(\vec{x}).  
\label{2dimlaplace}
\end{equation}

\noindent The lens mapping from the lens plane to the source plane, or
from $\vec{x}$ to $\vec{y}$, is a gradient mapping:

\begin{equation}
	\vec{y} 
   = \nabla\left(\frac12 \vec{x}\cdot\vec{x} - \psi(\vec{x}) \right),
\label{tllenseq}
\end{equation}

\noindent or,

\begin{equation}
	\vec{y} = \vec{x}-\nabla\psi(\vec{x}).  
\label{tllenseq2}
\end{equation}

\noindent Equation~(\ref{tllenseq}) is the lens equation of the
thin-lens approximation under the scaling defined by
Eqs.~(\ref{scaled}).

An alternative perspective is to derive the lens equation, Eq.~(\ref
{tllenseq}), by applying a stationarity condition to a Fermat
potential. If we define

\begin{equation}
	\widehat{\phi}(\vec{y},\vec{x}) 
   = \frac12(\vec{x} - \vec{y})^2 - \psi(\vec{x}),
\label{tlfermat}
\end{equation}

\noindent as a Fermat potential, the gradient mapping is equivalent to
the condition

\begin{equation}
	\nabla_x\widehat{\phi}(\vec{y},\vec{x}) = 0,
\end{equation}

\noindent i.e., the lens equation, Eq.~(\ref{tllenseq2}), follows from
Fermat's principle. (We generalize the idea of a Fermat potential in
Sec.~IV to a general lensing scenario, noting that other authors have 
also studied this problem extensively~\cite{Perlick0}.) Note that the 
second derivativeof $\hat{\phi}(\vec{y},\vec{x})$ with respect to 
$\vec{x}$ is equal tothe derivative of $\vec{y}$ with respect to 
$\vec{x}$ as defined inEq.~(\ref {tllenseq}).

The distortion of images can be studied via the Jacobian of the lens
mapping

\begin{equation}
	\mbox{\boldmath$A$}
   \equiv \frac{\partial\vec{y}}{\partial\vec{x}}
   =	  \frac{\partial^2\widehat{\phi}}
	       {\partial\vec{x}\partial\vec{x}}.
\end{equation}

\noindent We see that the Jacobian matrix is symmetric. This symmetry
exists, essentially, only in the special coordinates used. We will see
in Sec.~IV how this result arises non-perturbatively.

Using the lens equation, Eq.~(\ref{tllenseq}), we see that the Jacobian
matrix has the form

\begin{equation}
	\mbox{\boldmath$A$}
   =\left(\begin{array}{cc}
		1-\psi_{11} &  -\psi_{12} \\ 
		 -\psi_{21} & 1-\psi_{22}
	  \end{array}
    \right) ,  
\label{A1}
\end{equation}

\noindent where subscripts denote derivatives with respect to the
components of $\vec{x}=(x^1,x^2)$. By defining

\begin{eqnarray}
	    \gamma_1 (\vec{x}) 
   &\equiv& \frac12(\psi_{11} - \psi_{22}),  \nonumber \\
	    \gamma_2 (\vec{x}) 
   &\equiv& \psi_{12} = \psi_{21}  \label{gammadef}
\end{eqnarray}

\noindent and using Eq.~(\ref{2dimlaplace}), we can rewrite the
Jacobian matrix as

\begin{equation}
	\mbox{\boldmath$A$}
	=\left( 
		\begin{array}{cc}
		1-\kappa -\gamma_1 &          -\gamma_2 \\ 
			 -\gamma_2 & 1-\kappa +\gamma_1
		\end{array}
	 \right),  
\label{A2}
\end{equation}

\noindent where we have dropped the explicit $\vec{x}$ dependence in
$\kappa(\vec{x})$, $\gamma_1(\vec{x})$ and $\gamma_2(\vec{x})$.

It has become customary in the lensing community~\cite{EFS,BN} to refer
to $\kappa$ as the {\it convergence\/}, and $\gamma \equiv
\sqrt{\gamma_1^2+\gamma_2^2}$ as the {\it shear \/}. In this and the
subsequent paper~\cite{FKNII}, however, we will reserve the names of
convergence and shear for the optical scalars of a null geodesic
congruence. We show in our subsequent paper that (modulo terms
involving $D_s$ and $D_l$) $\kappa$ and $\gamma$ are {\it
essentially\/} the values of the convergence and shear of the
observer's past lightcone {\it at the lens plane\/} on the source's
side.

In terms of $\kappa$ and $\gamma$, the invariants of {\boldmath$A$} are

\begin{eqnarray}
   \det \mbox{\boldmath$A$} &=&(1-\kappa)^2-\gamma^2  \label{Adet} \\
   \mbox{tr \boldmath$A$}   &=&2(1-\kappa)  	      \label{Atr} \\
   a_{\pm }                 &=&1-\kappa \pm \gamma  \label{Aeigvalues}
\end{eqnarray}

\noindent where $a_{\pm }$ are the eigenvalues of the Jacobian matrix.
When $\det\mbox{\boldmath$A$}=0$, the lens mapping is singular and the
source lies at a caustic of the past cone. The quantity  $\mu=(\det
\mbox{\boldmath$A$})^{-1}$, referred to as the {\it magnification\/},
is the factor by which the image's brightness is magnified with respect
to the unlensed source.

Another important use of the Jacobian is to help understand the changes
in the shape of the image of a small source. To see this, we note that
a (suitably scaled) connecting vector, $\vec{X}$, at $\vec{x}$ in the lens
plane, is mapped into a (suitably scaled) connecting vector, $\vec{Y}$, at
$\vec{y}$ in the source plane, by {\boldmath$A$}, while 
$\mbox{\boldmath$A$}^{-1}$ maps source plane vectors back to the lens
plane:

\begin{equation}
   \vec{Y} = \mbox{\boldmath$A$} \vec{X}.  
\label{Amapping}
\end{equation}

\noindent For instance, if a small circle of radius $R$ at $\vec{y}$ in
the source plane is described by

\begin{equation}
 \vec{Y}(t)=\vec{y}+R\,\vec{s}(t)  
\label{tlcircle}
\end{equation}

\noindent where $\vec{s}(t)=(\cos t,\sin t)$, an image of this circle
is an ellipse, in the lens or image plane, described by

\begin{equation}
   \vec{X}(t)=\vec{x}+R\,\mbox{\boldmath$A$}^{-1}\vec{s}(t)  
\label{tlellipse}
\end{equation}

\noindent Hence, the Jacobian matrix, {\boldmath$A$}, and its inverse,
$\mbox{\boldmath$A$}^{-1}$, which is referred to as the {\it
magnification matrix\/}, play a large role in the description of the
images of small sources.

The semiaxes of the ellipse described by Eq.~(\ref{tlellipse}) must be
parallel to the principal axes of the magnification matrix
$\mbox{\boldmath$A$}^{-1}$, which are the same as the principal axes of
{\boldmath$A$}. This is because there are two vectors 
$\vec{s}(t_\pm)=(\cos t_\pm,\sin t_\pm)$,  which are parallel to the
two eigenvectors of {\boldmath$A$} and are mapped into themselves by
{\boldmath$A$} and $\mbox{\boldmath$A$}^{-1}$. These two directions
undergo the maximum stretching and squeezing because the
eigendirections of a symmetric linear map are those which have the
greatest and least stretching. Hence, the lengths of the major and
minor semiaxes are

\begin{equation}
	L_\pm = \frac{R}{|1 - \kappa \mp \gamma|}.
\end{equation}

The locations of these axes are given by the angles, $t_\pm$, which
satisfy

\begin{equation}
     \tan t_\pm  
   = \frac{\gamma_1}{\gamma_2}
	\mp \sqrt{\left(\frac{\gamma_1}{\gamma_2}\right)+1},  
\label{tlangles}
\end{equation}

\noindent i.e., in the directions

\begin{equation}
   \vec{X}(t_\pm)=\vec{x}+R\mbox{\boldmath$A$}^{-1}\vec{s}(t_\pm).
\end{equation}

\noindent This result is obtained [via extremizing
$(\vec{X}(t)-\vec{x})^2$ as a function of $t$] by finding the angles,
$t$, for vectors of the form of $\vec{s}(t)=(\cos t,\sin t)$ which are
the eigenvectors.

We are now in the position to define our shape parameters. The first
shape parameter is the ratio of the lengths of the semimajor and
semiminor axes, ${\cal R}$. This ratio is given by

\begin{equation}
    {\cal R}
   =\frac{L_+}{L_-}
   =\frac{|1-\kappa +\gamma |}{|1-\kappa -\gamma|}  
\label{tlratio}
\end{equation}

\noindent in the thin-lens approximation. The second shape parameter is
the area of the ellipse factored by $\pi$, or the product of the two
semiaxes, which we refer to as the area of the ellipse (since the area
is proportional to the product of the semiaxes with a factor of $\pi$).
For  the thin lens approximation, the (factored) area is given by

\begin{equation}
    	  {\cal A}
   \equiv L_+L_-
   =	  \frac{R^2}{(1-\kappa)^2-\gamma^2}
   =	  R^2\det (\mbox{\boldmath$A$}^{-1})  
\label{Area}
\end{equation}

\noindent The final shape parameter is the orientation of the major
semiaxis, $\delta$. By convention, we define $\delta$ to be $\delta
\equiv t_+$.

Thus, there are three quantities of interest in describing the image
of a circular source: the (suitably scaled) area of the image
{$\cal A$}, which encodes the overall scaling of the image compared
to the source; the ratio of the semiaxes $\cal R$ which is a measure of
the distortion up to scale; and the orientation of the semimajor axis
$\delta \equiv t_{+}$.  This definition of the orientation does not
have a definite meaning without a reference axis or when referred to a
circular source but it can be assigned a meaning for an elliptical
source when the discussion is generalized slightly as will be done
later in this article. We refer to these three quantities as the  {\it
shape parameters\/}. They will be defined non-perturbatively in the
next section.

\section{Non-perturbative image distortion}


We are interested in understanding image distortion within general
relativity with no approximations. We refer to this approach to image
distortion as a non-perturbative approach, to distinguish it from the
standard approach which is based on linearized perturbations around
flat space or an isotropic cosmological model.

To this effect, we need two things. First, we need a non-perturbative
counterpart of the lens mapping. Once the lens mapping is given, the
Jacobian matrix, or its inverse, the magnification matrix, can be
obtained. Second, we need non-perturbative definitions for image and
source shapes, and {\it distortion parameters\/}, which encode the
changes between the shapes of the source and the image, in terms of the
magnification matrix.

\subsection{Lens mapping}


At any fixed time the observer ``sees'' each source (on his or her past
lightcone) projected onto the observer's celestial sphere. By
introducing spherical coordinates, $(\theta,\phi)$, for each ray, via
this projection, the observer encodes a mapping between an image
location, $(\theta,\phi)$, and a source location. A formal lens mapping
can be obtained by the following considerations, summarized from
\cite{EFNspacetime,FNuniversal}. Similar considerations have been 
studied recently by other authors~\cite{Perlick1,Perlick2}. In particular,
see~\cite{blandfordkerr} for exact lensing in Kerr spacetime.

In a four dimensional Lorentzian space-time, $({\cal M},g_{ab}(z^a))$,
with local coordinates $z^a$, we assume that an expression for the past
lightcone of the observer, obtained by integrating the null geodesic
equations, is available. It has the form of four functions of three
parameters of the light cone at an observer's fixed time, $\tau$: 

\begin{equation}
   z^a=F^a(z_0^b(\tau),s,\theta,\phi).  
\label{pastlc}
\end{equation}

\noindent Here $z^a$ represents a point on the past lightcone of the
observer's worldline, $z_0^b(\tau)$ at time $\tau$, reached at an
affine length $s$ along the null geodesic with direction
$(\theta,\phi)$. The affine length $s$ has the value zero at the
observer and is scaled so that $g_{ab}\dot{F}^a\dot{z}_0^b=1$, using the 
notation $\dot{\;} \equiv \frac{\partial}{\partial s}$. Thefunctions $F^a$ 
can be obtained by integrating the geodesic equations,
$\ddot{F}^a+\Gamma_{bc}^a\dot{F}^b\dot{F}^c=0$ with the null condition

\begin{equation}
g_{ab}\dot{F}^a\dot{F}^b = 0.  
\label{nullcondition}
\end{equation}

\noindent The coordinates are arbitrary, but they can be chosen so that one
of them, say $z^0$, is timelike and the remaining three $z^i$ are
spacelike, i=1,2,3. In this case, for $a=0$, Eq.~(\ref{pastlc})
represents the time of departure of a light signal from the source to
the observer. On the other hand, Eq.~(\ref{pastlc}) for $a=1,2,3$ is
equivalent to a lens mapping between the angular position, $(z^2,z^3)$,
say, of a source of light and its image angles $(\theta,\phi)$ on the
observer's celestial sphere, plus a prescription of a radial
coordinate, $z^1$, which, in principle, can be obtained from
independent observations. Ideally, if the radial coordinate is known as
a function of the affine parameter $s$, then, in principle, the
parameter $s$ can be eliminated from $F^A(s,\theta,\phi)$, for $A=2,3$,
in terms of $(z^1,\theta,\phi)$, in which case a precise analog of the
standard lens mapping is obtained of the form
$z^A=G^A(z^1,\theta,\phi)$. For our purposes, it is not essential to
eliminate $s$, which may not even be possible in general (see
\cite{FKNschw} for the feasibility of such an elimination in the simple
case of a Schwarzschild spacetime). The non-perturbative lens mapping
$z^A=G^A(z^1,\theta,\phi)$ has been worked out explicitly in the case
of a Schwarzschild spacetime~\cite{FKNschw} and compared to standard
lensing.

{\it Aside.\/} Our non-perturbative version of a lens mapping differs
from the standard lens mapping, in addition to being non-perturbative,
in the sense that we do not use explicitly the Einstein equations -- we
simply consider any Lorentzian metric $g_{ab}(z^a)$ and its null
geodesics to construct the observer's past cones and the resulting lens
equation. In most thin-lens treatments of lensing the linearized
Einstein equations (in the form $\nabla^{2}\psi(\vec{x})
=2\kappa(\vec{x})$) are used explicitly from the start in the
construction of the lens equation.

\subsection{Jacobian of the Lens Mapping}

\begin{figure}
\centerline{\psfig{figure=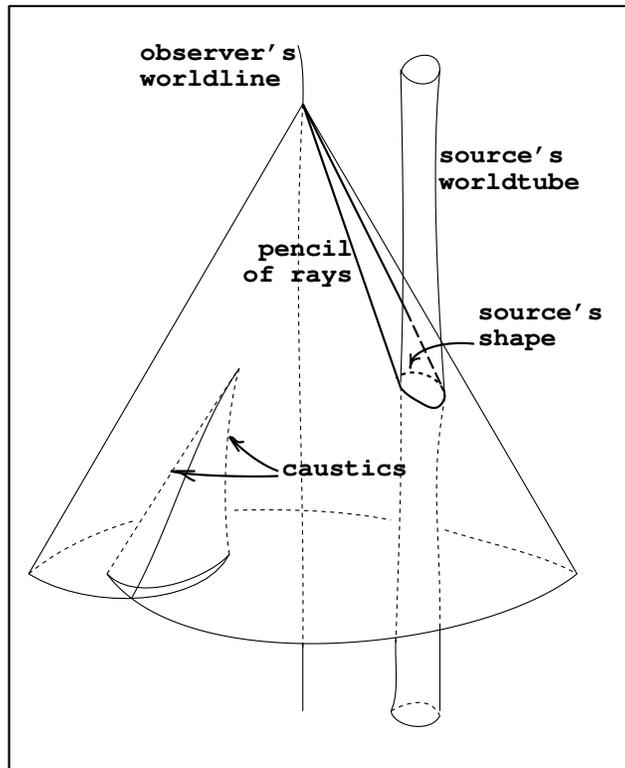,height=4in,angle=0}}

\caption{Diagram of image distortion.  The source's worltube intersects
the observer's past lightcone in a region free of caustics.  The
source's visible shape is defined by the intersection.  The pencil of
rays between the source's shape and the observer carries the shape of
the source into the shape of the image, on the observer's celestial
sphere. The pencil of rays can be described by geodesic deviation
vectors (connecting vectors of the observer's lightcone) from a central
null ray connecting the center of the source's shape to the observer.  }

\label{fig:distIIb}  
\end{figure}

Consider the situation where there is an extended source at some
distance from an observer.  The source's worldtube intersects the
observer's past lightcone, and we assume that the intersection occurs in
a region of the past lightcone where there are no caustics. See
Fig.~\ref{fig:distIIb}. The intersection determines the source's visible
shape. The pencil of rays that joins the shape of the source to the
observer ``carries'' the shape of the source into the shape of the
image, on the observer's celestial sphere. For the lensing purposes,
thus, an extended source consists of a collection of individual point
sources, all of which are connected to the observer by individual null
geodesics. In general, the image of an extended source must be obtained
by imaging each individual point via the non-perturbative lens equation.
However, if the sources are ``small,'' we can, in the spirit of the
previous Section, describe sources by {\it connecting vectors\/} along
one of the null geodesics (a ``central'' one), with image directions
$(\theta,\phi)$, that joins the source to the observer. Jacobi fields
(or connecting vectors) are solutions to the geodesic deviation equation
along that {\it fixed\/} geodesic. In the case of null geodesics, the
space of Jacobi fields is two-dimensional and can be obtained in our
approach directly from our non-perturbative lens mapping
Eq.~(\ref{pastlc}) in the following manner.

We note that there are three natural vectors that can be associated
with either the past lightcone or equivalently with the lens equation,
Eq.~(\ref{pastlc}). The first one is the tangent to the geodesics in
the lightcone:

\begin{equation}
   \ell^a \equiv \frac{\partial z^a}{\partial s} = \dot{F}^a.
\end{equation}

\noindent The remaining two are the Jacobi fields

\begin{equation}
   	  \widetilde{M}_1^a
   \equiv \frac{\partial z^a}{\partial\theta },
	  \hspace{1.5cm}
	  \widetilde{M}_2^a
   \equiv (\sin\theta)^{-1} \frac{\partial z^a}{\partial\phi}.  
\label{Ms}
\end{equation}

\noindent These two vectors are, in general, linearly independent and
connect neighboring geodesics at a fixed value of $s$, so they are two
linearly independent Jacobi fields (except at a caustic). They are
orthogonal to the tangent vector $\ell^a$ because they lie on the
lightcone: $g_{ab}\widetilde{M}_1^a\ell^b
=g_{ab}\widetilde{M}_2^a\ell^b =0$, and they are spacelike. Though any
other Jacobi field is a linear combination of
$(\widetilde{M}_1^a,\widetilde{M}_2^a)$ with constant parameters
$\alpha,\beta$,

\begin{equation}
Z^a=\alpha \widetilde{M}_1^a+\beta \widetilde{M}_2^a,
\end{equation}

\noindent we refer to $(\widetilde{M}_1^a,\widetilde{M}_2^a)$ as the
``natural'' Jacobi fields.

Two other important vector fields (not Jacobi fields) associated with
any null geodesic are the pair of spacelike, orthonormal, parallel
propagated vectors that are normal to $\ell^a$. They provide a means to
refer to directions that are ``the same'' at the orthogonal tangent
spaces at the source and at the observer. Analytically they are given
by

\begin{eqnarray}
	\ell^b\nabla_{\!b}e_1^a &=&\ell^b\nabla_{\!b}e_2^a=0, \\
	e_1\!\cdot \ell &=& e_2\!\cdot \ell =0, \\
	e_1\!\cdot e_1 &=& e_2\!\cdot e_2=-1, \\
	e_1\!\cdot e_2 &=&0,
\end{eqnarray}

\noindent where $v\!\cdot \!w$ denotes the scalar product between two
vectors, $g_{ab}v^aw^b$.

The natural connecting vectors,
$(\widetilde{M}_1^a,\widetilde{M}_2^a)$, in general, will have
transverse components along the space spanned by $(e_1^a,e_2^a)$, and a
longitudinal component along the geodesic congruence $\ell^a$. The
component along $\ell^a$ has no significance since it corresponds to
reparametrizations of the null geodesic. Jacobi fields  thus have two
essential transverse components. Two linearly independent Jacobi fields
can thus be thought as two two-component vectors that are obtained by
derivation from the lens mapping. Explicitly, this can be done by the
projection of $(\widetilde{M}_1^a,\widetilde{M}_2^a)$ into
$(e_1^a,e_2^a)$, thereby defining the matrix
$\widetilde{\mbox{\boldmath$J$}}(s)$

\begin{equation}
	(\widetilde{M}_1^a,\widetilde{M}_2^a)
   \rightarrow  \left[ \widetilde{\mbox{\boldmath$J$}}(s)
		\right] 
   \equiv \widetilde{\mbox{\boldmath$J$}}_i^j(s)
	= \widetilde{M}_i^ae_a^j.
\end{equation}

\noindent At the observer (i.e.,  at the apex of the cone), the
connecting vectors $(\widetilde{M}_1^a,\widetilde{M}_2^a)$ vanish, so
that $\widetilde{\mbox{\boldmath$J$}}(0) = 0$, and, furthermore, in the
limit as $s\rightarrow 0$ we have that

\begin{equation}
	\lim_{s\to 0}\frac{1}{s}
	\widetilde{\mbox{\boldmath$J$}}(s)
   =\mbox{\boldmath$I$}  
\label{lim}
\end{equation}

\noindent where {\boldmath$I$} is the identity matrix. This last result
is quite important and comes from the fact that although the connecting
vectors vanish at $s=0,$ their angular size has been chosen to be unity.
(At each value of $s$ close to the observer, each connecting vector 
subtends an arc from the observer's viewpoint. The connecting vector's 
arc is proportional to $s$, therefore the angular size of the
connecting  vector at the observer is well defined, and equal to the
derivative of  the connecting vector with respect to $s$.)

We adopt the following notation. When the matrix
$\widetilde{\mbox{\boldmath$J$}}(s)$ is evaluated at the source
position, denoted by $s=s^*$, it is called the {\it Jacobian\/} of the
lens mapping and is denoted by \mbox{\boldmath${}^*\!\!\widetilde{J}$}. 
In addition, we denote any quantity that
is defined or evaluated {\it at the source\/} by placing a $^*$ in
front of it, e.g.,
$({}^*\!\widetilde{M}_1^a,{}^*\!\widetilde{M}_2^a) =
(\widetilde{M}_1^a(s^*),\widetilde{M}_2^a(s^*))$.

Notice that, in a weak field-thin lens regime, our Jacobian 
\mbox{\boldmath${}^*\!\!\widetilde{J}$} is related to the Jacobian
\mbox{\boldmath$A$} of the thin-lens theory via

\begin{equation}
    \mbox{\boldmath$A$}
   =\frac{\mbox{\boldmath${}^*\!\!\widetilde{J}$}}{s^*}
\end{equation}

We now put a source at $s^*$ with an elliptical shape  and study how  it
is mapped back to the image space,the observer's celestial sphere. 
Though it is possible to do this construction with 
$(\widetilde{M}_1^a(s),\widetilde{M}_2^a(s))$ (or,equivalently, with 
$\widetilde{\mbox{\boldmath$J$}}_i^j(s)$) it turns out to be
considerably simpler to use a pair of Jacobi fields, $(M_1^a, M_2^a)$,
constructed from  a linear combination of the $\widetilde{M}_i^a$ that
are unit orthogonal  vectors at $s=s^*$, i.e., so that
$\mbox{\boldmath$J$}_i^j(s)\equiv  M_i^ae_a^j$ is the identity matrix at
$s=s^{*}$: 

\begin{equation}
	 \mbox{\boldmath$J$}_i^j(s^*)
  \equiv {}^*\!M_i^a{}^*\!e_a^j
  =	 \delta_i^j .
\end{equation}

\noindent It is easy to see that the linear combination must have the
form

\begin{equation}
	M_i^a
   =	(\mbox{\boldmath${}^*\!\!\widetilde{J}$}^{-1})_i^j\widetilde{M}_j^a ,
\end{equation}

\noindent so that 

\begin{equation}
	M_i^ae_a^k
   =	(\mbox{\boldmath${}^*\!\!\widetilde{J}$}^{-1})_i^j \widetilde{M}_j^ae_{a}^{k}
	\quad \Rightarrow 
	\mbox{\boldmath$J$}_i^k(s)
   =	(\mbox{\boldmath${}^*\!\!\widetilde{J}$}^{-1})_i^j
	\widetilde{\mbox{\boldmath$J$}}_j^k(s)  ,
\label{J}
\end{equation}

\noindent and hence
 
\begin{equation}
	\mbox{\boldmath$J$}_i^k(s^*)
  =	(\mbox{\boldmath${}^*\!\!\widetilde{J}$}^{-1})_i^j
	 \mbox{\boldmath${}^*\!\!\widetilde{J}$}_j^k
  =	\delta_i^k. 
\end{equation}

\noindent This is equivalent to assuming that we have chosen two
connecting vectors $M_1^a$ and $M_2^a$ so that

\begin{eqnarray}
{}^*\!M_1^a &=& {}^*\!e_1^a  \label{m1*} ,\\
{}^*\!M_2^a &=& {}^*\!e_2^a  \label{m2*} .
\end{eqnarray}

\begin{figure}
\centerline{\psfig{figure=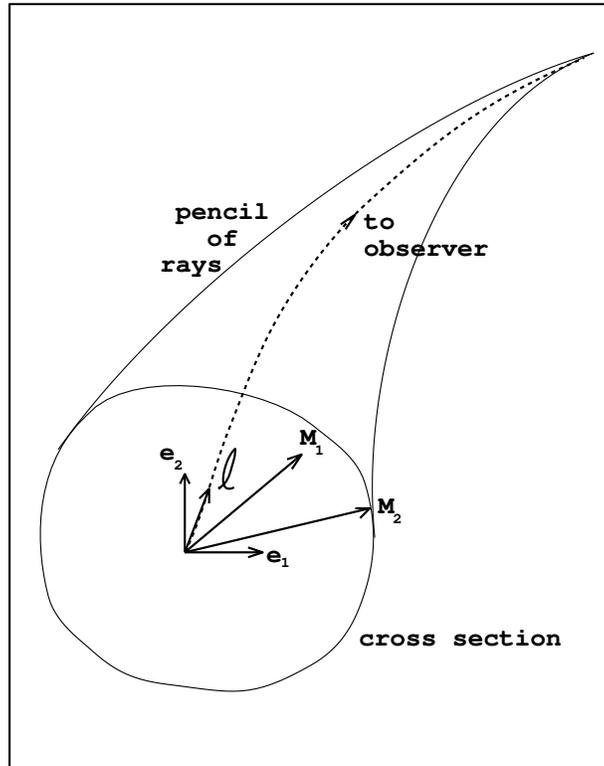,height=4in,angle=0}}

\caption{Vectors associated with the null geodesic that joins the source
to the observer.  The vector $\ell$ is tangent to the null geodesic. 
The vectors $(e_1,e_2)$ are spacelike, orthogonal to $\ell$, orthonormal
and parallel progagated along the geodesics. The vectors $(M_1,M_2)$ are
linearly independent Jacobi fields which coincide with $(e_1,e_2)$ at
the location of the source (not shown). }

\label{fig:distIIc}  
\end{figure}

\noindent In general, the two pairs $(M_{1}^{a},M_{2}^{a})$ and $%
(e_{1}^{a},e_{2}^{a})$, coincide at the point $s^{*}$, but not at any
other point along the null geodesic.  See Fig.~\ref{fig:distIIc}. 

For our construction in this article, it is a fundamental observation 
that, from Eqs.~(\ref{J}) and (\ref{lim}), as $s\to 0$ we have 

\begin{equation}
	\frac{1}{s}\mbox{\boldmath$J$}(s)
   \to  \mbox{\boldmath${}^*\!\!\widetilde{J}$}^{-1}, 
\end{equation}

\noindent i.e., in the limit as the null geodesic reaches the observer,
{\boldmath$J$} acts essentially as the inverse of the Jacobian of the
lens map, namely, as our {\it magnification matrix}.  We make extensive
use of this result throughout this and the companion
paper~\cite{FKNII}. 

Preliminarily to defining the shape and distortion parameters we point
out that even though the pairs $(e_1^a,e_2^a)$ and $(M_1^a,M_2^a)$
possess different meanings from each other, they can nevertheless be
expressed as linear combinations of each other. Assuming a relationship
of the form

\begin{eqnarray}
	e_1^a 
   &=&	 \alpha\cos(\lambda+\nu)M_1^a
        +\beta \cos \lambda     M_2^a,  \label{e1} \\
	e_2^a 
   &=&	 \alpha\sin(\lambda+\nu)M_1^a
	+\beta \sin \lambda     M_2^a,  \label{e2}
\end{eqnarray}

\noindent and using the properties of $e_i^a$ and its dual basis, namely

\begin{equation}
	e_i^ae_a^j=\delta_i^j,
	\qquad 
	e_i^a\ell_a=0,
	\qquad 
	\ell^b\nabla_{\!b}e_i^a=0 
\end{equation}

\noindent after a straightforward calculation the four parameters
$(\alpha,\beta,\nu,\lambda,)$ can be obtained in terms of the scalar
products $M_1\!\cdot M_1, M_2\!\cdot M_2$ and $M_1\!\cdot M_2$ as
follows:

\begin{eqnarray}
	\alpha &=&\sqrt{\frac{-M_2\!\cdot\!M_2}{C}},  \label{alpha} \\
	\beta  &=&\sqrt{\frac{-M_1\!\cdot\!M_1}{C}},  \label{beta} \\
      \cos \nu &=&\frac{M_1\!\cdot\!M_2}
	               {\sqrt{M_1\!\cdot\!M_1M_2\!\cdot\!M_2}},  
\label{nu}
\end{eqnarray}

\noindent where we use, for short, 

\begin{equation}
	 C
  \equiv M_{1}\!\cdot M_1M_2\!\cdot\!M_2-(M_1\!\cdot\!M_2)^2,
\end{equation}

\noindent and
 
\begin{equation}
     \dot{\lambda}
   = \frac{M_1\!\cdot\!M_2}{2\sqrt{C}}
	\frac{d}{ds}\log
	\left(\frac{M_2\!\cdot\!M_2}{M_1\!\cdot\!M_2}\right) .
\label{lambdadot}
\end{equation}

\noindent which fixes $\lambda $ up to a constant. The quantity $C$
represents the area spanned   by $(M_1^a,M_2^a)$, i.e.,  $C =
g^{ad}n_an_d$ with $n_a\equiv \epsilon_{abc}M_1^bM_2^c$.

Notice that the values of $(\alpha,\beta,\nu,\lambda )$ at $s^*$ are
determined to be

\begin{equation}
	{}^*\!\alpha =1,\hspace{1cm}
	{}^*\!\beta  =1,\hspace{1cm}
	{}^*\!\nu =-\frac{\pi}{2},\hspace{1cm}
	{}^*\!\lambda =\frac{\pi}{2}.
\end{equation}

\noindent We thus have the orthonormal parallel propagated basis
written in terms of the Jacobi fields.

\subsection{Shape and distortion parameters}


We now discuss the source parameters that enter into the considerations
of image distortion. Specifically we consider small elliptical sources,
which can be completely described by three parameters: the lengths of
the two semiaxes,  and the orientation of, say, the major semiaxis with
respect to a reference direction.  The two semiaxes can be equivalently
encoded into the ratio of major to minor semiaxis, and the product of
the two.  The product of the  two semiaxes is the area of the ellipse
divided by the number $\pi$.  In order to avoid carrying the factor
$\pi$ around, in this paper,  we simply refer to the product of the two
semiaxes as the area of the ellipse  and denote it by ${}^*\!\!{\cal
A}$.  We denote the ratio of the semiaxes by ${}^*\!{\cal R}$, and the
orientation of the major semiaxis with respect to the parallel
propagated direction $e_1^a$ by ${}^*\!\delta$. These are, by
definition, our {\it shape parameters\/} evaluated at the source, $s^*$.

At the source location, we have an orthonormal basis, 
$({}^*\!e_1^a,{}^{*}\!e_2^a)$, in terms of which an ellipse can be
expressed as

\begin{equation}
^{\ast }\!Y^{a}(t)\equiv
 Y^{1}(t)^{*}e_{1}^{a}+Y^{2}(t)^{*}e_{2}^{a}
=Y^{1}(t)^{*}M_{1}^{a}+Y^{2}(t)^{*}M_{2}^{a} 
\end{equation}

\noindent where the components $Y^i(t)$, $t$ being the curve parameter,
can be given by

\begin{eqnarray}
	\left(\begin{array}{c}
		Y^1 \\ 
		Y^2
		\end{array}
	\right) 
   &=&	\left(\begin{array}{cc}
		\cos\!{}^*\!\delta & -\sin\!{}^*\!\delta \\ 
		\sin\!{}^*\!\delta &  \cos\!{}^*\!\delta
		\end{array}
	\right) 
	\left(\begin{array}{cc}
		\sqrt{{}^*\!\!{\cal A}^*\!{\cal R}} & 0 \\ 
		0 & \sqrt{{}^*\!{\cal A}/{}^*\!{\cal R}}
		\end{array}
	\right) 
	\left(\begin{array}{c}
		\cos t \\ 
		\sin t
		\end{array}
	\right)  				\nonumber \\
   &\equiv &\sqrt{{}^*\!\!{\cal A}/{}^*\!{\cal R}}
	\left(\begin{array}{c}
	{}^*\!{\cal R}\cos t\cos\!{}^*\!\delta 
		     -\sin t\sin\!{}^*\!\delta \\ 
	{}^*\!{\cal R}\cos t\sin\!{}^*\!\delta 
		     +\sin t\cos\!{}^*\!\delta
		\end{array}
	\right)  \label{source}
\end{eqnarray}

\noindent As indicated, this arises from first taking a parametrized circle (the
vector $(\cos t,\sin t)$), then distorting into an ellipse of major
semiaxis $\sqrt{{}^*\!\!{\cal A}{}^*\!{\cal R}}$ (the product of the
semiaxes times the ratio of major to minor is the square of the major
axis) and minor semiaxis $\sqrt{{}^*\!{\cal A}/{}^*\!{\cal R}}$ (the
product of the semiaxes over the ratio of major to minor is the square
of the minor axis) and finally rotating through an arbitrary angle
${}^*\!\delta$.

{\emph{Remark:}} One should not confuse the curve parameter $t$ with the 
polar angle$\phi$ of each corresponding point on the curve relative to the
direction ${}^*\!e_1^a$. The polar angle $\varphi$ is related to $t$ at
the source by $\tan(\varphi-\delta_*)=\tan t/{}^*\!{\cal R}$.

In order to obtain the image of this source, we ``follow'' each point of
the ellipse back towards the observer using the Jacobi fields
$M_i^a(s)$, i.e., via

\begin{eqnarray}
	Z^a(s,t) 
   &\equiv &
	Y^1(t)M_1^a(s)+ Y^2(t)M_2^a(s)  \nonumber\\
   &=&
	\sqrt{\frac{{}^*\!\!{\cal A}}
	           {{}^*\!{\cal R}}}
	\Big(({}^*\!{\cal R}\cos t\cos\!{}^*\!\delta 
			   -\sin t\sin\!{}^*\!\delta) M_1^a
	    +({}^*\!{\cal R}\cos t\sin\!{}^*\!\delta
			   +\sin t\cos\!{}^*\!\delta) M_2^a
	\Big).
\label{image}
\end{eqnarray}

\noindent The vector field $Z^a$ is a Jacobi field for every value of
$t$, therefore it describes the image of the source near $s\to 0$, the
observer's location. On the other hand, it agrees with our expression
for the source at $s^*$, because of Eqs.~(\ref{m1*})-(\ref{m2*}).

Projecting the components of $Z^a(s,t)$ in the $e_i^a$ directions we
have

\begin{equation}
	Z^i(s,t)
   \equiv
	Z^a(s,t)e_a^i
   =	Y^j(t)M_j^a(s)e_a^i
   =	\mbox{\boldmath$J$}_j^i(s)Y^j(t); 
\label{image2}
\end{equation}

\noindent the matrix {\boldmath$J$} thus pulls the source's shape back
towards the observer. We then have the projected image of the source on
the observer's celestial sphere given by

\begin{equation}
	X^i(t)
   =	\lim_{s\to 0}\frac{1}{s}
	\mbox{\boldmath$J$}_j^i(s)Y^j(t)
   =	(\mbox{\boldmath${}^*\!\!\widetilde{J}$}^{-1})_j^iY^j(t) 
\end{equation}

\noindent or 
\begin{equation}
	X(t)=\mbox{\boldmath${}^*\!\!\widetilde{J}$}^{-1}Y(t). 
\end{equation}

\noindent This is the non-perturbative equivalent of
Eq.~(\ref{tlellipse}) of Section II. Notice, however, that we have not 
scaled the connecting vector at the source. (The connecting vector at
the observer must be scaled by $s$ in order for it not to vanish.)

The image $X^a(t)$ of the ellipse $Y^a(t)$ is an ellipse as well
because a non-singular continous linear transformation of an ellipse is
also an ellipse. In order to read off the image's elliptic parameters,
we start by extremizing the distance $Z\!\cdot Z$ as a function of the
parameter $t$. Directly from Eq.~(\ref{image}) we have

\begin{equation}
	Z\!\cdot Z
   = \frac{{}^*\!\!{\cal A}}{{}^*\!{\cal R}}
	  \left(a+b\sin^2t+c\sin t\cos t\right),
\end{equation}

\noindent with

\begin{eqnarray}
a(s) &=&{}^*\!{\cal R}^2
	\Big( (\cos\!{}^*\!\delta)^2 M_1\!\cdot\!M_1
	     +(\sin\!{}^*\!\delta)^2 M_2\!\cdot\!M_{2}
	     +2\sin\!{}^*\!\delta
	       \cos\!{}^*\!\delta M_1\!\cdot\!M_2
	\Big),  					\label{a} \\
b(s) &=&\frac12(1-\!{}^*\!{\cal R}^2)(M_1\!\cdot\!M_1+M_2\!\cdot\!M_2)
    +\frac12(1+\!{}^*\!{\cal R}^2)\cos(2{}^*\!\delta)
			        (M_2\!\cdot\!M_2-M_1\!\cdot\!M_1)
    -(1+\!{}^*\!{\cal R}^2)\sin(2{}^*\!\delta)     M_1\!\cdot\!M_2,  
							\label{b} \\
c(s) &=&{}^*\!{\cal R}
	\Big(\sin(2{}^*\!\delta)(M_2\!\cdot\!M_2-M_1\!\cdot\!M_1)
	    +2\cos(2{}^*\!\delta) M_1\!\cdot\!M_2
	\Big).  					\label{c}
\end{eqnarray}

\noindent Extremization ($d(Z\!\cdot\!Z)/dt=0$) yields

\begin{equation}
   \tan(2t) = -\frac{c}{b}.  
\label{extreme}
\end{equation}

\noindent There are, thus, four values of $t$ in the interval $[0,2\pi
]$ that extremize the distance to the center, and they differ by $\pi
/2$. Using Eq.~(\ref{extreme}), we can obtain, for $Z\!\cdot\!Z$,

\begin{equation}
	Z\!\cdot\!Z
   =\frac{{}^*\!\!{\cal A}}{2{}^*\!{\cal R}}
	\left( 2a+b\pm \sqrt{b^2+c^2}\right) .
\end{equation}

\noindent The lengths of the major and minor semiaxes, $L_+$ and $L_-$,
respectively, are thus

\begin{eqnarray}
	L_+ &=&\left(\frac{{}^*\!\!{\cal A}}{2{}^*\!{\cal R}}
		     \left( -2a-b+\sqrt{b^2+c^2}\right) 
	       \right)^{1/2} ,					\\
	L_- &=&\left(\frac{{}^*\!\!{\cal A}}{2{}^*\!{\cal R}}
		     \left( -2a-b-\sqrt{b^2+c^2}\right) 
	       \right)^{1/2}.
\end{eqnarray}

\noindent The ratio of the semiaxes is thus

\begin{equation}
	{\cal R} = \left(
			\frac{2a+b- \sqrt{b^2+c^2}}
			     {2a+b+ \sqrt{b^2+c^2}}
		   \right)^{1/2}
\label{ratio}
\end{equation}

\noindent where $a,b,c$ are explicit functions of the scalar products
between the two connecting vectors $M_1^a$ and $M_2^a$ given by
(\ref{a})-(\ref{c}).

The product of the semiaxes is what we refer to as the area of the
ellipse

\begin{equation}
     {\cal A}
   =
	\frac{{}^*\!\!{\cal A}} {2{}^*\!{\cal R}}
	\big(4a(a+b) - c^2\big)^{1/2}.
\end{equation}

\noindent With the parameters $a,b$ and $c$ given by Eq.~(\ref{a}),
Eq.~(\ref{b}) and Eq.~(\ref{c}) this becomes

\begin{equation}
     {\cal A}
   =
	{}^*\!\!{\cal A} \Big(M_1\!\cdot M_1 M_2\!\cdot M_2
				-(M_1\!\cdot M_2)^2\Big)^{1/2}			
\label{area}
\end{equation}

\noindent as is expected. 

In order to find the orientation $\delta$ of the ellipse with respect
to $e_1^a$ we need to express $Z^a$ in the basis $(e_1^a, e_2^a)$. This
is done by inverting Eqs.(\ref{e1})-(\ref{e2}), obtaining

\begin{eqnarray}
   M_1^a &= & \frac{  \cos\lambda e_2^a 
		    - \sin\lambda e_1^a}
		   { \alpha\sin\nu} 		,\label{M1}\\
   M_2^a &= & \frac{  \sin(\lambda+\nu)e_1^a 
 		    - \cos(\lambda+\nu)e_2^a}
		   {\beta\sin\nu} 		,\label{M2}
\end{eqnarray}

\noindent The orientation $\delta$ is then given by $\tan\delta \equiv
(Z\!\cdot e_2)/(Z\!\cdot e_1)$, or

\begin{equation}
     \tan\delta
   =
	\frac{ \alpha\cos(\lambda+\nu)
		({\cal R}_*\cos t\sin\delta_*+ \sin t\cos\delta_*)
	      -\beta\cos\lambda
	  	({\cal R}_*\cos t\cos\delta_* -\sin t\sin\delta_*)}
	     { \beta\sin\lambda
		({\cal R}_*\cos t\cos\delta_* -\sin t\sin\delta_*)
	      -\alpha\sin(\lambda+\nu)
		({\cal R}_*\cos t\sin\delta_*+ \sin t\cos\delta_*)}.
\label{delta}
\end{equation}

\noindent For generic values of $t$ this yields the polar angle that
the points on the ellipse make with the direction $e_1^a$.  The
orientation of the semiaxes is obtained by using, for $t$, the values
$t_\pm$ of  Eq.~(\ref{extreme}). It is straightforward to check that the
values of $t_\pm$ given by Eq.~(\ref{extreme}), which differ by $\pi/2$,
yield values of $\delta$ that also differ by $\pi/2$, as expected for
an ellipse. [Other authors~\cite{Panov} have studied the rotation of 
the image of an elliptical source but in a different physical context 
(the appearance of the polarization of radio emission by sources at
high redshift).]

We have thus obtained our shape parameters ${\cal R, A}$ and $\delta$
in terms of Jacobi fields, in Eq.~(\ref{ratio}), Eq.~(\ref{area}) and
Eq.~(\ref{delta}). 

In the limit as $s\to 0$ we can obtain, from these shape parameters,
the values of the elliptic parameters of the image.  Notice that both 
${\cal R}$ and $\delta$ are well defined and finite as $s\to 0$,
because they are ratios of connecting vectors that vanish at the same
rate (like $s$). Therefore, if we denote the ratio of semiaxes of the
image by ${\cal R}_I$, and their orientation by $\delta_I$, their
values are simply the values of ${\cal R}$ and $\delta$ at $s=0$:

\begin{eqnarray}
	{\cal R}_I &=& {\cal R}(0)	,\\
	\delta_I &= & \delta(0).
\end{eqnarray}

\noindent  The area of the image, however, is not a well-defined
concept, since the image lies on the celestial sphere of the observer,
which has no definite radius.  Furthermore, the shape parameter that
gives the area of the pencil or rays, ${\cal A}$, necessarily vanishes
at the observer's location (the apex of the lightcone).  Nevertheless,
the solid angle of the image (i. e., its ``angular'' area) is a
meaningful concept.  In order to obtain the image's solid angle, we
define the solid angle $\Omega(s)$ of the corresponding the pencil of
rays:

\begin{equation}
	\Omega(s) \equiv \frac {\cal A}{s^2}
\end{equation}

\noindent from which the image's solid angle $\Omega_I$ can be obtained by passing to
the limit as $s\to 0$, namely:

\begin{equation}
	\Omega_I = \lim_{s\to 0} \Omega(s).
\end{equation}

We define now the distortion of the image with respect to the source as
follows. We consider the image to be distorted if its semiaxes and
orientation are different from those of the source. Thus $(L_+/L_+^*),
(L_-/L_-^*)$, and $(\delta-\delta_*)$ are indicators of distortion. 
The image will be said to be distorted if any of the following {\it
distortion parameters\/} is nonvanishing

\begin{eqnarray}
	  {\frak D}_{\cal R} 
  &\equiv& \left(\frac{{\cal R}(0)}{{}^*\!{\cal R}}\right)^2 - 1
   =		\left(\frac{{\cal R}_I}{{}^*\!{\cal R}}\right)^2 - 1,\\
	  {\frak D}_{\Omega} 
  &\equiv& \lim_{s\to 0}
	   \left(
	   \frac{\Omega}{{}^*\Omega}
	   \,\right)^2					     - 1
   =	\left(
	   \frac{\Omega_I}{{}^*\Omega}
	   \,\right)^2					     - 1,\\
	  {\frak D}_{\delta} 
   &\equiv& \delta(0) - {}^*\!\delta
   =	\delta_I - {}^*\!\delta.
\end{eqnarray}	

\noindent These distortion parameters give a measure of ``total''
distortion through the gravitational field between a distant source and
the observer. The distortion accumulates, however, infinitesimally
along the null path.  It is the object of our subsequent paper
\cite{FKNII} to find the relationship between the ``infinitesimal
distortion'' through an infinitesimal displacement in the gravitational
field, and the optical scalars (convergence and shear) of the
lightcone. 

We conclude this section with an illustration of our approach to
distortion in the case of a Schwarzschild spacetime.


\subsection{Example: The Schwarzschild Case}

As an illustration, we consider now the case of gravitational lensing
by a Schwarzschild black hole of mass $m$. The line element is

\begin{equation}
    ds^2=\left(1-\frac{2m}{r}\right)dt^2
	-\left(1-\frac{2m}{r}\right)^{-1}\!\!dr^2 
	-r^2(d\theta^2+\sin^2\!\theta d\phi^2).
\label{line}
\end{equation}

\begin{figure}
\centerline{\psfig{figure=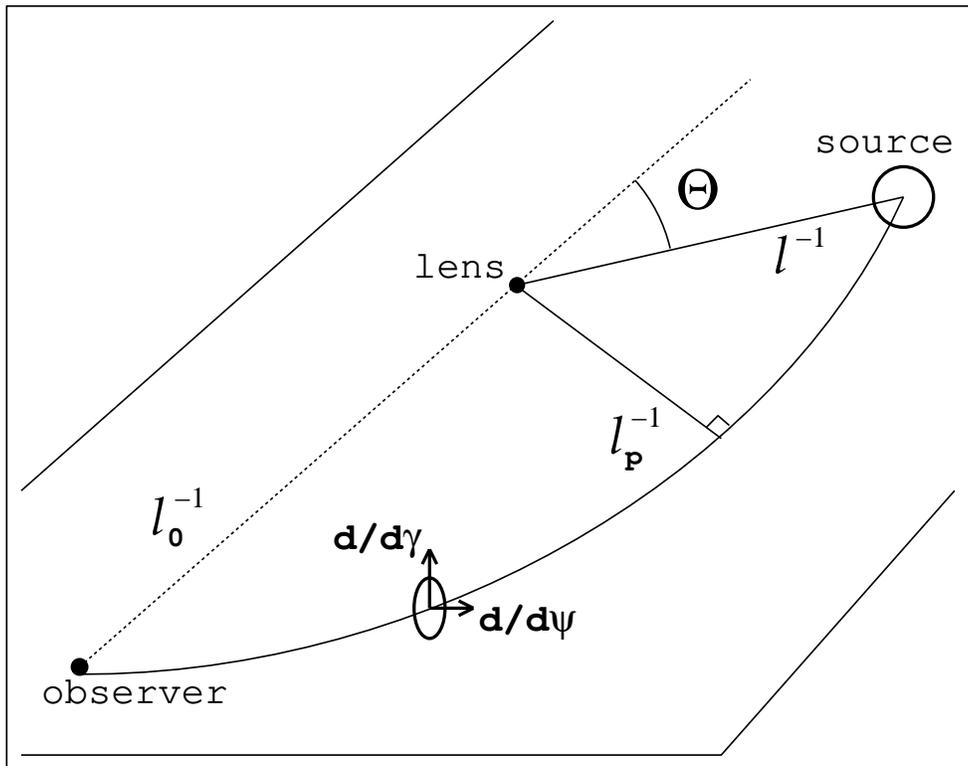,height=4in,angle=0} }

\caption{The connecting vectors in Schwarzschild spacetime.  The
vector $M_1\equiv\partial/\partial\psi$ lies on the plane defined by
the source, the observer and the lens.  The vector $M_2\equiv
\partial/\partial\gamma$ lies on the direction orthogonal to such a
plane. The inverse radial distance to the source
($l\equiv(\sqrt{2}r)^{-1}$) is used as a parameter along the null
geodesic that connects the source to the observer.  The function
$\Theta(l,\psi)$ represents the angular location of the source with
respect to the lens. }

\label{fig:distscheme}
\end{figure}

\noindent In this subsection only, we are using $(\theta,\phi)$ as the
regular  spherical Schwarzschild coordinates, instead of as coordinates
on the observer's celestial sphere. The null geodesics can be given
\cite{FKNschw,KNschw} in terms of two initial angles at the observer,
denoted $\psi$ and $\gamma$, representing the observer's celestial
sphere, and an inverse radial coordinate $l\equiv (\sqrt2 r)^{-1}$, in
place of an affine parameter $s$:

\begin{eqnarray}\label{t}
   t
   & =&
	t_0
       +2\int_{l_0}^{l_p}		    
	\frac{\sqrt{l_p^2(1-2\sqrt{2}ml_p)}}
	     {\sqrt{l_p^2(1-2\sqrt{2}ml_p)-l^2(1-2\sqrt{2}ml)}}
		\frac{dl'}{\sqrt{2}l'^2(1-2\sqrt{2}ml')}\nonumber\\
   &&
	 + \int_{l}^{l_0}		   
	\frac{\sqrt{l_p^2(1-2\sqrt{2}ml_p)}}
	     {\sqrt{l_p^2(1-2\sqrt{2}ml_p)-l^2(1-2\sqrt{2}ml)}}
	\frac{dl'}{\sqrt{2}l'^2(1-2\sqrt{2}ml')} 		,\\
	l
   &=&  l							,\\ 
	\cos\theta
   &=&
	-\cos\theta_0\cos\Theta
	+\sin\theta_0\sin\Theta\cos\gamma	,\label{theta}\\
	\tan\phi
   &=&
	\frac{
	 \sin\phi_0\sin\theta_0
	-\tan\Theta
	 \left(
	 	 \cos\phi_0\sin\gamma
		-\sin\phi_0\cos\gamma\cos\theta_0
	 \right)
		}
	      {
	 \cos\phi_0\sin\theta_0
	+\tan\Theta
	\left(
		 \sin\phi_0\sin\gamma
		+\cos\phi_0\cos\gamma\cos\theta_0
	\right)
		}			,\label{phi}
\end{eqnarray}

\noindent where $(t_0,l_0,\theta_0,\phi_0)$ are the coordinates of the
observer's location, $l_p$ is the inverse radial distance of closest
approach, defined as the smallest of the positive roots of 

\begin{equation}
   \sin^2\!\!\psi\, l_p^2(1-2\sqrt{2}ml_p) 
		  - l_0^2(1-2\sqrt{2}ml_0) = 0,
\end{equation}

\noindent and where $\Theta(l,l_0,\psi)$ is the following function:

\begin{eqnarray}\label{Theta}
  \Theta (l, l_0, \psi)
     &=&
	  \pi 
	- 2\int_{l_0}^{l_p} 	
	   \frac{dl}
		{\sqrt{l_p^2(1-2\sqrt{2}ml_p)-l^2(1-2\sqrt{2}ml)}} 
	-\int_{l}^{l_0} 	
	 \frac{dl'}
		{\sqrt{l_p^2(1-2\sqrt{2}ml_p)-l'^2(1-2\sqrt{2}ml')}},	
\end{eqnarray}

\noindent for positive values of $\psi$.  For negative values of $\psi$
we have 

\begin{equation}
	\Theta(l,l_0,\psi)=-\Theta(l,l_0,-\psi),
\end{equation}

\noindent namely, $\Theta$ is an odd function of $\psi$. Equations
(\ref{t})-(\ref{phi}) describe points on the past lightcone of the
observer which lie at an affine distance $s$ larger than the affine
distance of closest approach, so they can be interpreted as source
points behind the lens.  Eqs.~(\ref{t})-(\ref{phi}) are the equivalent
of Eq.~(\ref{pastlc}), and Eqs.~(\ref{theta})-(\ref{phi}) represent strictly the
lens mapping between the image's angular location $(\psi,\gamma)$ and
the source's angular location $(\theta,\phi)$.  By derivation of the
lens mapping we define the starting connecting vectors

\begin{eqnarray}
   \widehat{M}_1^a
    &\equiv& 
	\left(\frac{\partial t}{\partial \psi},
	 \frac{\partial l}{\partial \psi},
	 \frac{\partial \theta}{\partial \psi},
	 \frac{\partial \phi}{\partial \psi}
	 \right)
    =
	\left(\frac{\partial t}{\partial \psi},
	 0,
	 \frac{\partial \theta}{\partial \psi},
	 \frac{\partial \phi}{\partial \psi}
	 \right),				\\
   \widehat{M}_2^a
    &\equiv& 
	\left(\frac{\partial t}{\partial \gamma},
	 \frac{\partial l}{\partial \gamma},
	 \frac{\partial \theta}{\partial \gamma},
	 \frac{\partial \phi}{\partial \gamma}
	 \right)
    =
	\left(0,
	 0,
	 \frac{\partial \theta}{\partial \gamma},
	 \frac{\partial \phi}{\partial \gamma}
	 \right),						
\end{eqnarray}

\noindent where $t,\theta,\phi$ are functions of $(l,\psi,\gamma)$
given by Eqs.(\ref{t}), (\ref{theta}), (\ref{phi}), and the partial
derivatives are taken at fixed value of $l$. See Figure
\ref{fig:distscheme}. $(\widehat{M}_1^a, \widehat{M}_2^a)$ are actually
different from the ``natural'' connecting vectors introduced in Section
III in that  $(\widehat{M}_1^a, \widehat{M}_2^a)$ are {\it not\/}
normalized in a manner that would ensure their angular size to be 1. We
have

\begin{eqnarray}
\frac{\partial \theta}{\partial \gamma}
    &=&
	\frac{\sin\Theta\sin\theta_0\sin\gamma	 }
	     {\sqrt{ 1 -(\cos\Theta\cos\theta_0
			 -\sin\Theta\sin\theta_0\cos\gamma)^2
		    }
              }	,				\label{connectinga}
\\
\frac{\partial \theta}{\partial \psi}
    &=&
	-\frac{	 \sin\Theta\cos\theta_0
		+\cos\Theta\sin\theta_0\cos\gamma}
	     {\sqrt{ 1 -(\cos\Theta\cos\theta_0
			 -\sin\Theta\sin\theta_0\cos\gamma)^2
		    }
              }							
	\frac{\partial \Theta}{\partial \psi},	\label{connectingb}
\\
 \frac{\partial \phi}{\partial \gamma}
    &=&
	-\frac{\sin\Theta( \sin\Theta\cos\theta_0
			 +\cos\Theta\sin\theta_0\cos\gamma)}
	     {1-(\cos\Theta\cos\theta_0
			 -\sin\Theta\sin\theta_0\cos\gamma)^2},
						\label{connectingc}\\
 \frac{\partial \phi}{\partial \psi}
    &=&
	-\frac{\sin\theta_0\sin\gamma}
	     {1-(\cos\Theta\cos\theta_0
			 -\sin\Theta\sin\theta_0\cos\gamma)^2}	
	\frac{\partial \Theta}{\partial \psi}.	\label{connectingd}
\end{eqnarray}

\noindent Notice that, by Eq.~(\ref{connectinga}) and Eq.~(\ref{connectingc}),
the vector $\widehat{M}_2^a$ is proportional to $\sin\Theta$ for all
generic values of $\theta_0$, except for $\theta_0=0, \pi$.  This means
that, generically, the vector $\widehat{M}_2^a$ vanishes at $\Theta=0$,
which, by Eqs.~(\ref{theta})-(\ref{phi}), represents source points
$(\theta,\phi)$ along the optical axis. These are the caustics of the
past lightcone of the observer. With the metric Eq.~(\ref{line}) we have  

\begin{eqnarray}
	\widehat{M}_1\!\cdot\! \widehat{M}_1 
  & =& 
	-\frac{1}{2l^2}\left(1-\frac{l^2(1-2\sqrt{2}ml)}
				    {l_p^2(1-2\sqrt{2}ml_p)}
			\right)
	\left(\frac{\partial\Theta}{\partial \psi}\right)^2  ,  
\label{m11} \\
	\widehat{M}_2\!\cdot\! \widehat{M}_2 
  & =& 
	 -\frac{\sin^2\Theta}{2l^2}			,\label{m22}\\
	\widehat{M}_1\!\cdot\! \widehat{M}_2
  & =&
	0.
\end{eqnarray}

\noindent Notice that these two connecting vectors are orthogonal along
the lightcone. This is a feature of the spherical symmetry of the
spacetime, which induces the axial symmetry of the lightcone. No two
other Jacobi fields in general will be orthogonal, so these two are a
special basis of Jacobi fields. However, they don't have unit length. 
In accordance with the previous subsection, we define two connecting
vectors now that are orthonormal at the source's location:

\begin{equation}
	M_1^a \equiv \frac{\widehat{M}_1^a}			 
{\sqrt{-{}^*\!(\widehat{M}_1\!\cdot\!\widehat{M}_1)}},
	\hspace{1cm}
	M_2^a \equiv \frac{\widehat{M}_2^a}
{\sqrt{-{}^*\!(\widehat{M}_2\!\cdot\!\widehat{M}_2)}}.
\end{equation}

In our example, we will first examine a circular source, which we can
express by

\begin{equation}
	Z^a(l,t) = \cos t\, M_1^a(l)
		+\sin t \,M_2^a(l).
\end{equation}

\noindent Because $M_1^a$ and $M_2^a$ remain orthogonal, one can see
that the circle at $l_*$ is distorted to an ellipse at any other point
$l$, with semiaxes 

\begin{eqnarray}
	L_+ &=& \sqrt{-M_2\!\cdot\! M_2}
	     =
		\sqrt{\frac{\widehat{M}_2\!\cdot\! \widehat{M}_2}
		   {{}^*\!(\widehat{M}_2\!\cdot\! \widehat{M}_2)}},
\\
	L_- &=& \sqrt{-M_1\!\cdot\! M_1}
	     =
		\sqrt{\frac{\widehat{M}_1\!\cdot\! \widehat{M}_1}
		   {{}^*\!(\widehat{M}_1\!\cdot\! \widehat{M}_1)}},
\end{eqnarray}

\noindent and the ratio of the semiaxes of the ellipse at the
observer's location is given simply by  

\begin{equation}
 {\cal R}= 	\sqrt{\frac{{}^0\!(M_2\!\cdot\! M_2)}
			    {{}^0\!(M_1\!\cdot\! M_1)}
			}
         =
	\sqrt{\frac{{}^*\!(\widehat{M}_1\!\cdot\!\widehat{M}_1)}
		   {{}^*\!(\widehat{M}_2\!\cdot\!\widehat{M}_2)}
	      \frac{{}^0\!(\widehat{M}_2\!\cdot\!\widehat{M}_2)}
		   {{}^0\!(\widehat{M}_1\!\cdot\!\widehat{M}_1)}
			},
\end{equation}

\noindent where we use the notation ${}^0\!(v\!\cdot\! w) \equiv
\!v\!\cdot\! w |_{l_0}$.  With Eqs.~(\ref{m11})-(\ref{m22}) we obtain 

\begin{equation}\label{Ratio}
{\cal R} = \left|
		\frac{\sin\Theta(l_0)}
		     {\partial\Theta/\partial\psi|_{l_0}}
		\frac{\partial\Theta/\partial\psi|_{l_*}}
		     {\sin\Theta(l_*)}
		\sqrt{
		\frac{l_p^2(1-2\sqrt{2}ml_p)-l_*^2(1-2\sqrt{2}ml_*)}
		     {l_p^2(1-2\sqrt{2}ml_p)-l_0^2(1-2\sqrt{2}ml_0)}
		     }
		\right|.
\end{equation}

\noindent The factor
$\frac{\sin\Theta(l_0)}{\partial\Theta/\partial\psi|_{l_0}}$ represents
the ratio of the norms of the two starting connecting vectors at the
observer's location (the apex of the lightcone), at which all
connecting vectors vanish. However the ratio of the norms is finite and
can be calculated in closed form via a limiting procedure~\cite{FKNschw}, 
obtaining

\begin{equation}
	\frac{\sin\Theta(l_0)}{\partial\Theta/\partial\psi|_{l_0}}
=
	-2\sin\psi\cos\psi.
\end{equation}

\begin{figure}
\centerline{\psfig{figure=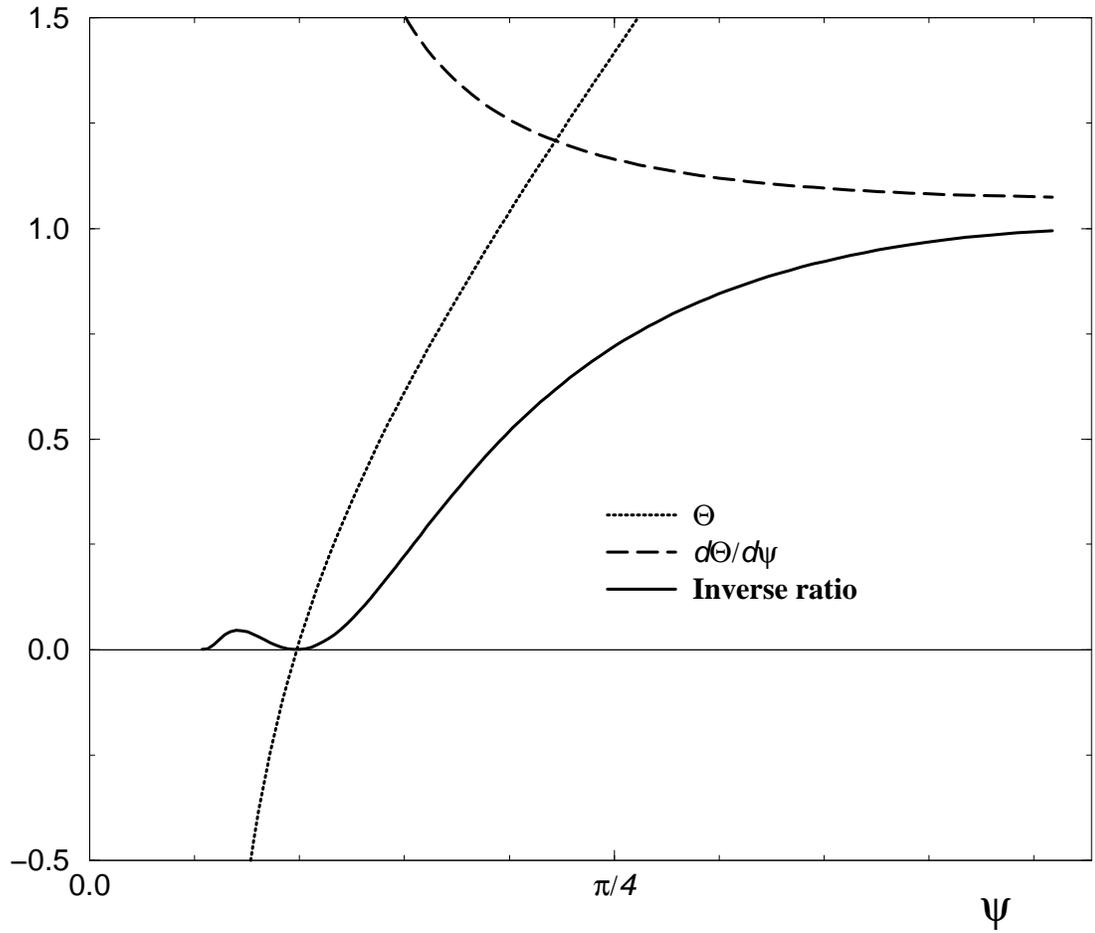,height=5in,angle=-90}}

\caption{Distortion by a Schwarzschild black hole. The inverse ratio
of the semiaxes of the image divided by the inverse ratio of the
semiaxes of the source is plotted for fixed distance between the
source and the deflector, equal to the distance between the lens and
the observer, as a function of the image angle.  For large image
angle the distortion decreases, since the ratio tends to 1 as
$\psi\to\pi/2$. For small image angle, the geodesics that arrive at
the observer go through a strong gravitational field, and the
distortion is large.  The distortion is infinite (where the inverse
ratio is zero) for the images of sources that lie on the caustics
the observer's lightcone.  }

\label{fig:schwarschildratio}
\end{figure}

\noindent The other factors in the ratio Eq.~(\ref{Ratio}) must be
evaluated numerically. Figure~\ref{fig:schwarschildratio} shows a plot
of the inverse ratio ${\cal R}^{-1}$ at fixed value of the inverse
radial distance to the source $l_*$, as the image angle varies.  We can
see that the inverse ratio, representing the minor axis over the major
axis, approaches 1 for large image angles (lightrays that pass far from
the lens), but is everywhere smaller than 1, which means that the
images are elongated along the polar direction around the optical axis
in the lens plane (see Fig. \ref{fig:distscheme}). The inverse ratio
vanishes at the caustics, namely, when the source lies along the
optical axis, behind the lens. This means that the distortion increases
dramatically when the source is close to a caustic, leading to the
formation of arcs.  We have chosen to plot the inverse ratio rather
than ${\cal R}$ itself, because the major axis becomes infinite for
sources lying at the caustics, which makes ${\cal R}$ less convenient
to plot.

\begin{figure}
\centerline{\psfig{figure=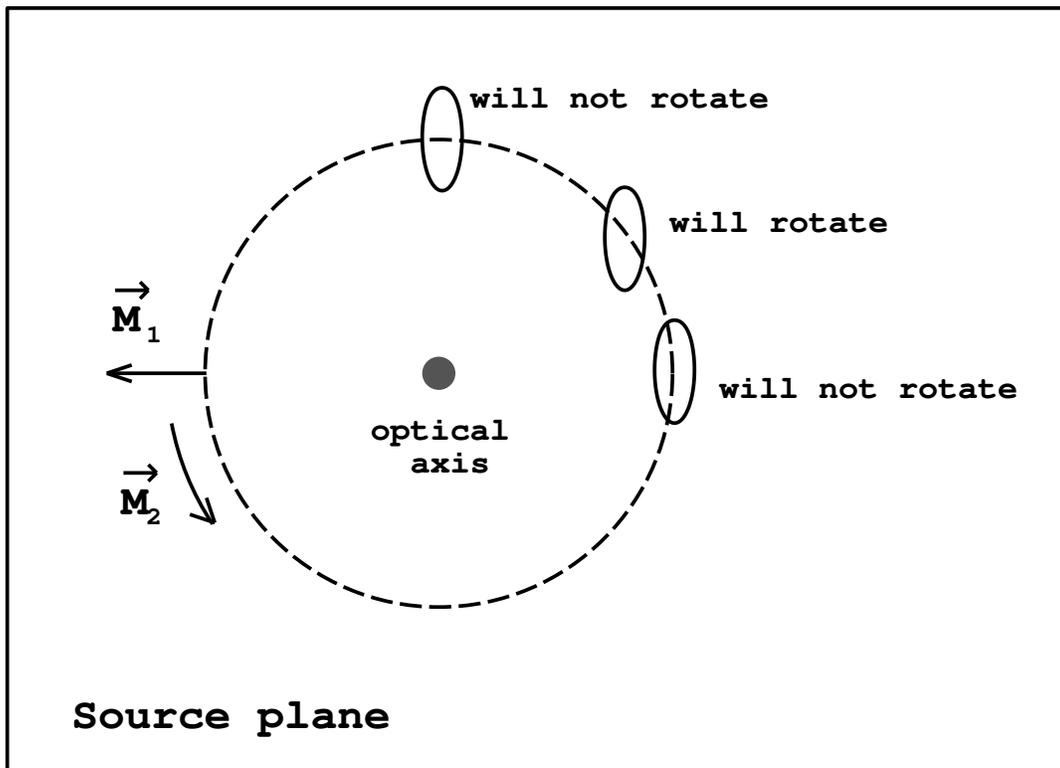,height=4in,angle=0}}

\caption{Distortion by a Schwarzschild black hole. The axial
symmetry in the source plane leads to no orientation distortion if
the semiaxes of the elliptical source are aligned with the
connecting vectors $M_1^a$ and $M_2^a$. $M_1^a$ points in the polar
angular direction, whereas $M_2^a$ points along the radial direction
on the source plane. Thus, only sources aligned with the radial
direction experience no distortion of orientation.}
\label{fig:distortionI3} 
\end{figure}

In the general case of the images of elliptical sources, they are given
by the construction Eq.~(\ref{image2}). In our case

\begin{equation}
	\left(\begin{array}{c}
		\!Z^1\!\!\\
		\!Z^2\!\!
	      \end{array}
	\right) = \left(\begin{array}{cc}
			\sqrt{-M_1\!\cdot\! M_1}&0\\
			0&\sqrt{-M_2\!\cdot\! M_2}
		    \end{array}
		\right)
		\left(\begin{array}{c}
				\!Y^1\!\!\\
				\!Y^2\!\!
			\end{array}
		\right)
		= \mbox{\boldmath $J$}\left(\begin{array}{c}
				\!Y^1\!\!\\
				\!Y^2\!\!
			\end{array}
		\right)
\end{equation}

\noindent which shows, in the limit as $l\to l_0$,  that the
magnification matrix, and therefore, the Jacobian matrix as well, in
this case is symmetric, as is in the case of standard gravitational
lensing. The stretching  of the elliptical source occurs then along
the eigendirections of {\boldmath $J$}, which are the directions of
$M_1^a$ and $M_2^a$, since the matrix is diagonal. This means that if
the source's semiaxes are aligned with the vectors $(M_1^a,M_2^a)$
then the image will not be rotated with respect to the source, so the
distortion parameter measuring orientation distortion will vanish,
${\frak D}_\delta =0$. However, if the source's semiaxes are not
aligned with $(M_1^a,M_2^a)$, then the image will be rotated, and
${\frak D}_\delta$ will not vanish. See Fig.~\ref{fig:distortionI3}

\section{Non-perturbative use of Fermat's principle in lensing
}\label{sec:gauge}


In this section, we want to give a view of distortion that is based
on the existence of families of null surfaces - this view in turn
leads to a non-perturbative version of Fermat's principle applied to
lensing.  Although to apply it in practice is difficult, there  being
so few cases where enough null surfaces are known, it can be
implemented perturbatively.  In any  case, it yields a slightly
different picture of what is taking place in lensing.

We begin by assuming that a two-point function,
$G(x^{a},x_{0}^{a},\zeta ,\overline{\zeta}),$ is known where for
\textit{each} value of the parameters ($\zeta ,\overline{\zeta}),$
$G=0$, represents a \textit{null surface} containing the two points
$(x^{a},x_{0}^{a}).$ One should think of $x_{0}^{a}$ as being fixed
and $G=0$ as representing a two-parameter family of null surfaces
containing $x_{0}^{a};$ the parameters could be thought of as
representing the sphere of null directions at $x_{0}^{a}.$ The
gradient of $G,$ i.e., $\partial _{a}G\equiv G_{a}\equiv \ell_{a},$
is a null covector, $g^{ab}G_{a}G_{b}=0$.

Though we make no use of it now, we mention, for later use, that
there is a special choice of a two point function that is often quite
useful; if we begin with an asymptotically flat spacetime with the
past null boundary, $S^{2} \times R$, referred to as $\frak{I}^{-},$
and coordinatized by ($\zeta ,\overline{\zeta })$ on the $S^{2}$ and
$u$ on the $R,$ then the future lightcone of each point $(u,\zeta
,\overline{\zeta })$ of $\frak{I}^{-}$ can be written as
$u=Z(x^{a},\zeta ,\overline{\zeta }).$ Since by definition the
gradient of $Z,$ i.e, $\partial _{a}Z$ is a null covector, then by
defining

\begin{equation} G(x^{a},x_{0}^{a},\zeta ,\overline{\zeta })\equiv
Z(x^{a},\zeta
,\overline{\zeta })-Z(x_{0}^{a},\zeta ,\overline{\zeta }), \label{2pt}
\end{equation}

\noindent we have the two point function satisfying our null surface
conditions.

We want to construct the envelope of $G(x^{a},x_{0}^{a},\zeta
,\overline{\zeta })=0$ under variations of ($\zeta ,\overline{\zeta
})$.  The envelope is a special null surface, namely the lightcone
through $x_{0}^{a}$.  Explicitly, this is done by setting to zero the
($\zeta ,\overline{\zeta })$ derivatives of $G(x^{a},x_{0}^{a},\zeta
,\overline{\zeta }).$ The envelope is then given by:

\begin{eqnarray} G(x^{a},x_{0}^{a},\zeta ,\overline{\zeta }) &=&0,
\nonumber \\
\partial
_{\zeta }G(x^{a},x_{0}^{a},\zeta ,\overline{\zeta }) &=&0, \nonumber \\
\partial
_{\overline{\zeta }}G(x^{a},x_{0}^{a},\zeta ,\overline{\zeta }) &=&0. 
\label{Envelope}
\end{eqnarray}

\noindent If the second two equations \textit{could} be solved for
$(\zeta ,\overline{\zeta })=(\Gamma
(x^{a},x_{0}^{a}),\overline{\Gamma }(x^{a},x_{0}^{a})$, they could be
substituted into $G(x^{a},x_{0}^{a},\zeta ,\overline{\zeta })=0,$
yielding $g(x^{a},x_{0}^{a})\equiv G(x^{a},x_{0}^{a},\Gamma
,\overline{\Gamma })=0$, the analytic expression for the lightcone of
$x_0^a$. Whether or not they could be solved, the three equations,
Eqs.~(\ref{Envelope}), could be solved for some three $x^{i}$ of the
four $x^{a}$ in terms of the fourth coordinate $x^{*}$ and the
parameters, i.e.,

\begin{equation} x^{i}=x^{i}(x^{*},x_{0}^{a},\zeta,\overline{\zeta}). 
\label{lghtcn}
\end{equation}

\noindent Equation~(\ref{lghtcn}) is the parametric form of the
lightcone of $x_{0}^{a}$, where the parameters, $(\zeta ,\overline{\zeta
})$, label the generators of the lightcone ($(\zeta,\bar\zeta)$ and play the
same role as $(\theta,\phi)$ did in previous sections.)

It is from this parametric form of the lightcone that we will construct
the Jacobian matrix of the lens equation.  Note that if the $ x^{*}=r $
is a radial type of coordinate, then $x^{a}(r,x_{0}^{a},\zeta ,
\overline{\zeta })=\{x^{i}=x^{i}(r,x_{0}^{a},\zeta ,\overline{\zeta })$,
$ x^{*}=r\}$ form the lens and time of arrival equations where the
observer is at $x_{0}^{a}(\tau)$.  The derivatives of $x^{a}$ with
respect to $(r,\zeta ,\overline{\zeta })$ are respectively the null
geodesic tangent vector and the geodesic deviation vectors.

We first point out that the construction we just gave can be given an
alternative meaning; if the equation $G(x^{a},x_{0}^{a},\zeta
,\overline{ \zeta })=0$ is solved for the coordinate $t=x^{0},$ so that

\[ G=0\Leftrightarrow t=\widehat{G}(x^{\alpha
},x_{0}^{a},,\overline{\zeta }), 
\]

\noindent then the envelope construction is identical to
``extremizing'' the time, $t$, with respect to ($\zeta
,\overline{\zeta})$ variations.  We thus have constructed the
lightcone of $x_{0}^{a}$ via Fermat's principle.  It is then clear
that the construction of the lens equation that was given in Sec.~II
via the Fermat potential is a special case of the more general
construction we have just outlined.  It will be seen that from this
construction there are natural coordinates so that the Jacobian
matrix is symmetric.

By treating the three equations, Eqs.~(\ref{Envelope}) and
$x^{*}=r=constant$ as four \textit{implicit} equations for the
determination of $x^{a}$ as functions of ($\zeta ,\overline{\zeta }$)
we can, by differentiation with respect to $(\zeta
,\overline{\zeta})$ construct the deviation vectors, $M^{a}=\partial
_{\zeta }x^{a}$ and $\overline{M}^{a}=\partial _{\overline{\zeta
}}x^{a}$. Explicitly, the derivative  of Eqs.~(\ref{Envelope}) and
$x^{*}=r$ with respect to $\zeta$ are

\begin{eqnarray} M^{a}G_{a}+\partial _{\zeta }G(x^{a},x_{0}^{a},\zeta
,\overline{\zeta
}) =M^{a}G_{a}&=&0, \nonumber \\
M^{a}\partial_{\zeta}G_{a}(x^{a},x_{0}^{a},\zeta
,\overline{\zeta } )+\partial_{\zeta}\partial_{\zeta
}G(x^{a},x_{0}^{a},\zeta , \overline{\zeta }) &=&0, \nonumber \\
M^{a}\partial
_{\overline{\zeta }}G_{a}(x^{a},x_{0}^{a},\zeta ,\overline{ \zeta
})+\partial
_{\zeta
}\partial _{\overline{\zeta }}G(x^{a},x_{0}^{a}, \zeta ,\overline{\zeta
})
&=&0,
\nonumber \\ M^{a}\delta _{a}^{*} \equiv (\partial _{\zeta }x^{*}) &=&0,
\label{dev1}
\end{eqnarray}

\noindent and the complex conjugate equations are

\begin{eqnarray} \overline{M}^{a}G_{a}+\partial _{\overline{\zeta
}}G(x^{a},x_{0}^{a},\zeta , \overline{\zeta })
=\overline{M}^{a}G_{a}&=&0,\nonumber \\
\overline{M}^{a}\partial _{\zeta }G_{a}(x^{a},x_{0}^{a},\zeta ,\overline{
\zeta
})+\partial _{\overline{\zeta }}\partial _{\zeta }G(x^{a},x_{0}^{a},\zeta
,\overline{\zeta }) &=&0, \nonumber \\ \overline{M}^{a}\partial
_{\overline{\zeta
}}G_{a}(x^{a},x_{0}^{a},\zeta , \overline{\zeta })+\partial
_{\overline{\zeta
}}\partial
_{\overline{\zeta } }G(x^{a},x_{0}^{a},\zeta ,\overline{\zeta }) &=&0,
\nonumber \\
\overline{M}^{a}\delta _{a}^{*} &=&0. \label{dev1*} \end{eqnarray}

\noindent In the first of each of the two sets we simplified using the
envelope condition.  We thus see that there are only two non-vanishing
tetrad components for both $M^{a}$ and $\overline{M}^{a}$. We define
these components as

\begin{eqnarray*} M^{+} &\equiv &M^{a}\partial _{\zeta
}G_{a}(x^{a},x_{0}^{a},\zeta
,\overline{ \zeta })=-\partial _{\zeta }\partial _{\zeta
}G(x^{a},x_{0}^{a},\zeta ,
\overline{\zeta }), \\ M^{-} &\equiv &M^{a}\partial _{\overline{\zeta
}}G_{a}(x^{a},x_{0}^{a},\zeta ,\overline{\zeta })=-\partial
_{\overline{\zeta
}}\partial
_{\zeta }G(x^{a},x_{0}^{a},\zeta ,\overline{\zeta }), \\ \overline{M}^{+}
&\equiv
&\overline{M}^{a}\partial _{\zeta }G_{a}(x^{a},x_{0}^{a},\zeta
,\overline{\zeta
})=-\partial _{\overline{\zeta }}\partial _{\zeta
}G(x^{a},x_{0}^{a},\zeta
,\overline{\zeta }), \\ \overline{M}^{-} &\equiv
&\overline{M}^{a}\partial
_{\overline{\zeta } }G_{a}(x^{a},x_{0}^{a},\zeta ,\overline{\zeta
})=-\partial
_{\overline{\zeta }}\partial _{\overline{\zeta }}G(x^{a},x_{0}^{a},\zeta
,\overline{\zeta }), \end{eqnarray*}

\noindent and we have the tetrad version $\frak J$ of the Jacobian
matrix,$\widetilde{\mbox{\boldmath$J$}}$, i.e.,

\begin{equation} \frak{J}\equiv \left(\begin{array}{cc} M^{+} &
\overline{M}^{+} \\
M^{-} & \overline{M}^{-} \end{array} \right) = \left( \begin{array}{cc}
-\partial
_{\zeta }\partial _{\zeta }G & -\partial _{\overline{\zeta } }\partial
_{\zeta
}G \\
-\partial _{\overline{\zeta }}\partial _{\zeta }G & -\partial
_{\overline{
\zeta
}}\partial _{\overline{\zeta }}G \end{array} \right). \label{tetA}
\end{equation}

\noindent (Note that $\partial_\zeta G_a$ and
$\partial_{\bar\zeta}G_a$ play the same role as did $e^i_a$ in
previous sections). We see that the Jacobian matrix is symmetric when
expressed in the tetrad  basis.  A local coordinate system (depending
on the value of ($\zeta ,\overline{\zeta })$ or equivalently on the
relevant null geodesic between source and observer) can always be
introduced so that the tetrad components of $\frak{J}$ are also the
coordinate components.

It should be noted that though there is no easy way to relate
$\frak{J}$ directly to the shear and divergence of the lightcone
congruence, there nevertheless is a rather startling relationship
between $\frak{J}$ and certain asymptotic quantities (asymptotic
shears) in the case of asymptotically flat spacetimes when the
two-point null surface functions $G$ are given by Eq.~(\ref{2pt}). 
This relationship is both startling and (at least so-far) rather
incomprehensible.  At the moment it does not seem to have any bearing
on astrophysically measurable quantitites and is only of theoretical
interest.

Using Eq.~(\ref{2pt}), Eq.~(\ref{tetA}) becomes

\[ {\frak J}(x^{a},x_{0}^{a},\zeta ,\overline{\zeta })= \left(
\begin{array}{cc}
-[\partial _{\zeta }\partial _{\zeta }Z-\partial _{\zeta }\partial
_{\zeta
}Z_{0}] &
-[\partial _{\overline{\zeta }}\partial _{\zeta }Z-\partial _{
\overline{\zeta
}}\partial _{\zeta }Z_{0}] \\ -[\partial _{\overline{\zeta }}\partial
_{\zeta
}Z-\partial _{\overline{ \zeta }}\partial _{\zeta }Z_{0}] & -[\partial
_{\overline{\zeta
}}\partial _{ \overline{\zeta }}Z-\partial _{\overline{\zeta }}\partial
_{\overline{\zeta } }Z_{0}] \end{array} \right) .  \]

\noindent Though, as we mentioned earlier, the function
$u=Z(x^{a},\zeta ,\overline{ \zeta })$ represents the future lightcone
from the point $(u,\zeta ,\overline{ \zeta  })$ of $\frak{I}^{-}$ , it
has an alternative reciprocal meaning:  If $x^{a}$ is held constant and
$(\zeta ,\overline{\zeta })$ are varied over the sphere, then
$u=Z(x^{a},\zeta,\overline{\zeta})$ describes the the intersection of
the past lightcone of the point $x^{a}$ with $\frak{I}^{-},$ the past
lightcone cut of the point $x^{a}$.  This past lightcone, as it
approaches in the limit $\frak{I}^{-},$ has an asymptotic shear $\sigma
_{x}(x^{a},\zeta ,\overline{\zeta })$

It has been shown in the completely different context of the Null
Surface Reformulation of general relativity \cite{nsf1,nsf2} that
${\frak J}(x^{a},x_{0}^{a},\zeta , \overline{\zeta })$ has an
asymptotic interpretation, namely

\[ {\frak J}(x^{a},x_{0}^{a},\zeta ,\overline{\zeta })=
\left(\begin{array}{cc}
-[\sigma
_{x}(x^{a},\zeta ,\overline{\zeta })-\sigma _{x_{0}}(x_{0}^{a},\zeta
,\overline{\zeta
})] & -[R-R_{0}] \\ -[R-R_{0}] & -[\overline{\sigma }_{x}(x^{a},\zeta
,\overline{\zeta
})- \overline{\sigma }_{x_{0}}(x_{0}^{a},\zeta ,\overline{\zeta })]
\end{array}
\right)
\]

\noindent where $R$ and $R_{0}$ are, respectively, measures of the
curvature of each of the lightcone cuts from $x^{a}$ and $x_{0}^{a}.$
This matrix constructed from the differences between the two
asymptotic shears are surprisingly a measure of the changes in
geodesic deviation of geodesics going from $x^{a}$ and $x_{0}^{a}.$
It is this fact that we find both startling and incomprehensible.

\section{Conclusions}


We have developed a non-perturbative description of distortion of
images to be used in a generic spacetime in a manner analogous 
to the case of standard gravitational
lensing, where a set of simplifying assumptions is normally used. 

We have introduced three shape parameters which describe a small,
elliptically shaped cross-section of a pencil of rays from an
``elliptical'' source in the past lightcone of an observer. These are
the cross-sectional area ${\cal A}$ of the pencil (or its angular
analog $\Omega$); the ratio of the major to minor semiaxes of the
ellipse, ${\cal R}$; and the orientation of the ellipse, $\delta$
with respect to some arbitrary reference axis. By propagating these
quantities towards the observer, we have found the shape parameters
of the image, which lies on the observer's celestial sphere. Based on
the comparison of the shape parameters of the source and image, we
have defined three distortion parameters, the values of which provide
a quantitative measure of the severity of the distortion introduced
by the gravitational field on the path of the pencil of rays.

Our work in this paper is based on the use of Jacobi fields to
propagate the source along the lightcone.  Our approach can be
directly related to astrophysical gravitational lensing in the 
weak-field regime where the thin lens approximation is used by 
noticing that, within such a regime, 
the Jacobian matrix \mbox{\boldmath$A$} is equal to our Jacobian 
matrix \mbox{\boldmath$^*\!\!\widetilde{J}$} scaled by the distance to
the source, $s^*$, namely:

\[
	\mbox{\boldmath$A$}
   =	\frac{\mbox{\boldmath$^*\!\!\widetilde{J}$}}{s^*}.
\]

\noindent When the weak-field thin-lens regime is not imposed, however,
there is no geometric meaning to $\mbox{\boldmath$\widetilde{J}$}/s$.
This is a reason why we prefer to refer to 
$\mbox{\boldmath$\widetilde{J}$}$ itself as the Jacobian and consider
it as the closest meaningful analog of the matrix \mbox{\boldmath$A$}
in use in astrophysical gravitational lensing. 

Although our calculations in this paper do reproduce the formalism of
image distortion of standard lensing theory in the regimes of weak
fields, thin lenses and small angles, they do not make any
assumptions about the strength of the gravitational field or about
the thickness of the mass distribution.  Interesting prospective
applications of this approach include the development of alternative
approximation techniques to the lensing problem in which the
gravitational field is weak (so that the metric can be found from the
linearized Einstein equations, or equivalently, by adding
Schwarzschild-like contributions) but where one utilizes a fully 
three-dimensional mass distribution instead of multiple lens
planes. 

Having defined the shape parameters of the pencil of rays in terms of 
the Jacobi fields of the observer's lightcone, our next goal is to
clarify the relationship between the optical scalars of the lightcone
and the change in the shape parameters of the pencil of rays. We derive
this relationship in our companion paper \cite{FKNII}.  In addition, in
our companion paper~\cite{FKNII} we derive the expression of the
Jacobian matrix \mbox{\boldmath$A$} in terms of the optical scalars of
the lightcone in the weak-field thin-lens approximation.


\acknowledgments

We are indebted to Volker Perlick for kindly pointing out to us the
reference~\cite{Panov}, and to J\"{u}rgen Ehlers for stimulating
conversation. This work was supported by the NSF under grants  No. PHY
98-03301, PHY92-05109 and PHY 97-22049.



\newpage 

\newpage

\newpage

\newpage 

\newpage

\newpage

\end{document}